\documentclass[aps,prl,twocolumn,superscriptaddress]{revtex4-1}

\newcommand {\etal}{{\it et} {\it al.}}
\newcommand {\rxx}{\rho _{xx}}
\newcommand {\ryx}{\rho _{yx}}

\newcommand {\sxy}{\sigma _{xy}}

\newcommand {\pio}{{\rm Pr}_{2}{\rm Ir}_{2}{\rm O}_{7}}

\usepackage[dvipdfmx]{graphicx}
\usepackage{bm}
\usepackage{pifont}
\usepackage{amsmath,amssymb}
\usepackage[dvipdfmx]{color}

\begin{document}

\title{Experimental signatures of versatile Weyl semimetal in pyrochlore iridate with spin-ice like magnetic orders}

\author{Kentaro Ueda}
\affiliation{Department of Applied Physics and Quantum Phase Electronics Center (QPEC), University of Tokyo, Tokyo 113-8656, Japan}

\author{Hiroaki Ishizuka}
\affiliation{Department of Physics, Tokyo Institute of Technology, Tokyo 152-8551, Japan}

\author{Markus Kriener}
\affiliation{RIKEN Center for Emergent Matter Science (CEMS), RIKEN Advanced Science Institute (ASI), Wako 351-0198, Japan}

\author{Shunsuke Kitou}
\affiliation{RIKEN Center for Emergent Matter Science (CEMS), RIKEN Advanced Science Institute (ASI), Wako 351-0198, Japan}

\author{Denis Maryenko}
\affiliation{RIKEN Center for Emergent Matter Science (CEMS), RIKEN Advanced Science Institute (ASI), Wako 351-0198, Japan}

\author{Minoru Kawamura}
\affiliation{RIKEN Center for Emergent Matter Science (CEMS), RIKEN Advanced Science Institute (ASI), Wako 351-0198, Japan}

\author{Taka-hisa Arima}
\affiliation{Department of Applied Physics and Quantum Phase Electronics Center (QPEC), University of Tokyo, Tokyo 113-8656, Japan}
\affiliation{RIKEN Center for Emergent Matter Science (CEMS), RIKEN Advanced Science Institute (ASI), Wako 351-0198, Japan}

\author{Masashi Kawasaki}
\affiliation{Department of Applied Physics and Quantum Phase Electronics Center (QPEC), University of Tokyo, Tokyo 113-8656, Japan}
\affiliation{RIKEN Center for Emergent Matter Science (CEMS), RIKEN Advanced Science Institute (ASI), Wako 351-0198, Japan}

\author{Yoshinori Tokura}
\affiliation{Department of Applied Physics and Quantum Phase Electronics Center (QPEC), University of Tokyo, Tokyo 113-8656, Japan}
\affiliation{RIKEN Center for Emergent Matter Science (CEMS), RIKEN Advanced Science Institute (ASI), Wako 351-0198, Japan}
\affiliation{Tokyo College, University of Tokyo, Tokyo 113-8656, Japan}

\date{\today}

\begin{abstract}
We report experimental signatures of topological transitions among the Weyl semimetal states of pyrochlore $\pio $, where the Kondo coupling between the Ir topological electrons and the spin-ice like orders of Pr moments plays a decisive role. 
The magnetic-field dependence of resistivity and the Hall conductivity exhibits a plateau and a sharp jump associated with a magnetic-field hysteresis, similar to a liquid-gas-like transition in dipolar spin ice system.
Furthermore, the Kondo coupling is controlled by the hydrostatic pressure, revealing that the field-induced displacement of Weyl points in the momentum space strongly depends on the respective electronic state as well as on the Kondo coupling strength.
These observations pave a route toward the engineering of band topology in hybrid quantum materials with relativistic conduction electrons and localized magnetic moments.
\end{abstract}


\maketitle

\noindent

Geometrical frustration, which gives rise to unusual magnetic ground states and excitations, is a long-standing issue in condensed matter physics.
A well-studied example is pyrochlore magnets, composed of a three-dimensional network of corner-sharing tetrahedra (Fig. 1(a)) \cite{2010RMPGardner}.
For instance, pyrochlore rare-earth ($R$) titanates turn into the disordered spin ice (SI) state under the constraint that two of four spins on the vertices of the tetrahedron point inward to its center and the other two outward (2-in 2-out (2/2) configuration).
This local constraint named `ice rule' after analogous proton bond disorder in water ice produces interesting properties such as the Pauling's residual entropy \cite{1999NatureRamires,2001ScienceBramwell} and fractional excitations \cite{2008NatureCastelnovo,2009ScienceFennell}.
One advantage of a magnetic system in frustration physics is to manipulate the zero-point entropy by an external magnetic field.
In particular, the application of the field along the [111] crystalline direction, which is parallel to the anisotropic axis of the apical spin on the (111) triangular layer (Fig. 1(a)), firstly polarizes the apical spin to confine the 2/2 disorder into the two-dimensional kagom$\rm \acute{e}$ planes, i.e. kagom$\rm \acute{e}$ ice (KI), followed by the fully-polarized 3-in 1-out (3/1) state in a sufficiently large field \cite{1998PRLHarris,2003PRLSakakibara,2003JPSJHiroi}.

Recent intensive studies on topological electronic states have spurred renewed interest in frustrated systems \cite{2014NCommMazin,2015PRLBergholtz,2018NatureYe,2020NatureYin}.
Among them, particular attention has been focused on the family of pyrochlore iridates $R_2$Ir$_2$O$_7$ ($R$ being rare-earth ions or Y) as promising magnetic topological materials; $R_2$Ir$_2$O$_7$ is the first system proposed to be an antiferromagnetic Weyl semimetal (WSM) with an all-in all-out order of the Ir moments \cite{2011PRBWan}, and furthermore the interplay between localized $R$-$4f$ moments and Ir conduction electrons is expected to produce a variety of WSM states \cite{2018PRXYao,2018PRBOh}.
Among the $R_2$Ir$_2$O$_7$ family, $\pio $ and Nd$_2$Ir$_2$O$_7$ are located near the metal-insulator boundary. In paramagnetic phase, their electronic structures are characterized by the quadratic band touching (QBT) at the $\Gamma$ ($k$=0) point that is protected by cubic symmetry \cite{2013PRLMoon,2015NCommKondo}.
Importantly, the symmetry breaking is expected to yield topologically-nontrivial electronic states;
in fact, $\pio $ and Nd$_2$Ir$_2$O$_7$ have been found to exhibit a variety of interesting properties including unusual anomalous Hall effect \cite{2010NatureMachida,2018NCommUeda}, magnetic field-induced metal-insulator transitions \cite{2015PRLUeda,2016NPhysTian,2017NCommUeda}, and anomalous metallic states on magnetic domain walls \cite{2014PRBUeda,2015ScienceMa}. These observations imply that near the metal-insulator boundary the magnetic field stabilizes a variety of rare-earth magnetic configurations, and more importantly, broadens the region of the topological electronic phases with Weyl nodes. However, the magnetic ground state of $\pio $ remains controversial, in part due to the difficulty in keeping the right material stoichiometry inherent to volatile starting materials \cite{2015PRBMacLaughlin}.

In this work, we carefully prepared and characterized single-crystalline $\pio $ (see Ref. \cite{SM}) and performed thorough investigations by means of magnetization and specific-heat measurement, as well as the electrical transport down to 0.05 K and under hydrostatic pressure for the close examination of the ground state.
The precise single-crystal structure analysis reveals the anti-site mixing for several samples.
For the close-stoichiometric sample, we find that Pr $4f$ moments undergo a long-range magnetic order with SI manifold at zero field and a cascade of field-induced metamagnetic transitions to KI and 3/1 states, akin to the dipolar spin-ice system. Both longitudinal and Hall resistivities show significant changes across the transitions, indicative of the topological Ir $5d$ electron band modulation via the $f$-$d$ Kondo coupling in the respective WSM states.  To possibly control the $f$-$d$ Kondo coupling, we exploit the application of high pressure at the base temperature 0.05 K and observe a remarkable (minimal) pressure effect on the Hall conductivity in the 3/1 (KI) state.
Our theoretical calculation reveals that the field-induced displacement of Weyl points in the momentum space strongly depends on the respective WSM state as well as on the Kondo coupling strength, leading to the significant variation of Hall effect under pressures as observed.

A single crystal of Pr$_2$Ir$_2$O$_7$ was grown by KF flux method \cite{2007MRBMillican}.
We find the strong sample dependence of magnetotransport properties, which is presumably correlated to the local structural imperfections (see details in Ref. \cite{SM}.)
We chose the nearly-stoichiometric sample that reaches the largest magnetoresistance ratio of $\sim $50 \% at 2 K and 14 T (see Fig. S1(d)), which is much larger than that in the previous report \cite{2011PRLBalicas}.
Magnetization and transport measurements above 0.39 K were performed by using MPMS and PPMS (Quantum Design equipped with $^3$He refrigerators).
The sample was cooled down to 0.05 K of a dilution refrigerator, combined with a Be-Cu pressure cell which allows us to apply the pressure up to 1.5 GPa.

\begin{figure}
\begin{center}
\includegraphics[width=2.5in,keepaspectratio=true]{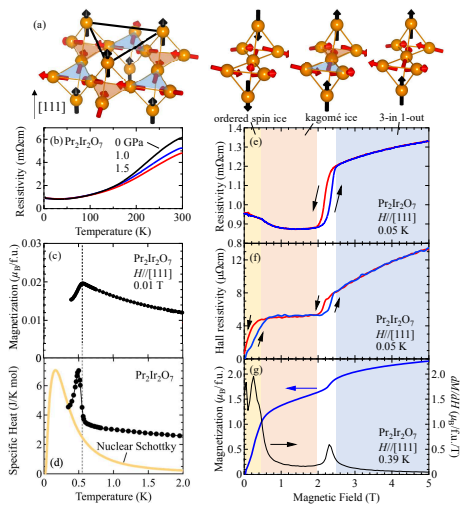}
\caption{(color online).
(a) Pyrochlore lattice composed of the kagom$\rm \acute{e}$ layer and triangular layer stacking along [111] crystalline direction.
(b) Temperature dependence of resistivity at several pressures.
Temperature dependence of (c) magnetization at 0.01 T and (d) specific heat below 2 K. The thick yellow line denotes the nuclear Schottky anomaly.
Magnetic field dependence of (e) resistivity at 0.05 K, (f) Hall resistivity at 0.05 K, and (g) magnetization and its field derivative at 0.39 K along [111] crystalline direction. Schematic pictures of magnetic configurations are displayed upon the panels.
}
\end{center}
\end{figure}

Figure 1(b) displays the temperature dependence of resistivity $\rxx $ under several hydrostatic pressures. $\rxx $ exhibits metallic-like temperature dependence except for a small upturn at around 50 K, as reported previously \cite{2006PRLNakatsuji}.
The application of pressure slightly decreases $\rxx $ at high temperatures. Figures 1(c) and 1(d) show the temperature dependence of magnetization and specific heat below 2 K, respectively. While the magnetization exhibits a clear cusp at $T_{\rm c}$=0.6 K, the specific heat peaks up sharply at 0.5 K. This small but clear discrepancy in temperatures of these anomalies indicates the spin freezing which thus hampers the long-range ordering at $T_{\rm c}$, vide infra \cite{2013NPhysPomaranski}.
The first-order-like-transition feature in specific heat was also reported in stuffed polycrystalline Pr$_{2+x}$Ir$_{2-x}$O$_{7-\delta }$ \cite{2015PRBMacLaughlin} which reveals the magnetic Bragg peak with the propagation wave vector $q_{m}$=[001] below $T_{\rm c}$=0.93 K.
Remarkably, the observed first-order-like transition with $q_{m}$=[001] spin configuration is consistent with the theoretical study by Melko et al. \cite{2001PRLMelko}, although their model is based on ferromagnetic dipolar interaction.
Such a magnetic order is hereafter termed `ordered SI', which is different from the disordered SI in titanates.
It should be noted that the previously reported $T_{ \rm c}$ is somewhat higher than our result, which can imply that the observed magnetic order is very sensitive to the stoichiometry, or equivalently to the carrier density in the Ir electron band.
It indicates that the Ruderman-Kittel-Kasuya-Yoshida (RKKY) interaction mediated by Ir conduction electrons, which is by one order of magnitude larger than the direct exchange interactions between Pr moments \cite{2006PRLNakatsuji}, plays a key role in the observed magnetic order, as theoretically demonstrated \cite{2014JPSCPIshizuka}.
This suggestion is supported by the absence of long-range order down to 0.02 K in the non-$d$-electron counterparts Pr$_2$Zr$_2$O$_7$ \cite{2013NCommKimura,2017PRLWen}.

\begin{figure}
\begin{center}
\includegraphics[width=2.5in,keepaspectratio=true]{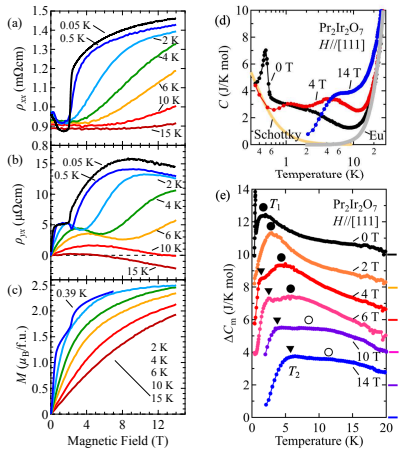}
\caption{(color online).
Magnetic field dependence of (a) resistivity, (b) Hall resistivity, and (c) magnetization at several temperatures.
(d) Temperature dependence of specific heat at 0 T (black markers), 4 T (red markers), and 14 T (blue markers). The thick yellow line denotes the nuclear Schottkey anomaly and the gray line denotes the specific heat of Eu$_2$Ir$_2$O$_7$ representing the lattice contribution.
(e) The magnetic contribution of specific heat $\Delta C_{\rm m}$ as a function of temperature at several magnetic fields. Each data is shifted vertically by 2 K/Kmol for clarity. The circles and triangles indicate the maxima of hump structures defined as temperature $T_{1}$ and $T_{2}$, respectively.
Because of the broad features at high fields, open circles are plotted for $T_{1}$.
}
\end{center}
\end{figure}

An important aspect of SI is the series of metamagnetic transitions induced by an external magnetic field $H$, especially when it is applied along the [111] direction. Figures 1(e)-(g) show $H$ dependence of $\rxx $ and Hall resistivity $\ryx $ at the lowest temperature in this measurement, 0.05 K, and the magnetization curve at 0.39 K for $H$//[111].
As $H$ increases, $\rxx $ decreases slightly, then stays almost constant between 0.6 and 2 T, and abruptly increases by $\sim $40 \% at $\mu _{0}H_{c}$=2.5 T, accompanied by a clear hysteresis between increasing and decreasing $H$. The observed magnetoresistance shows quite a distinct feature in both magnitude and $H$-dependence from the previously reported result \cite{2011PRLBalicas}.
$\ryx $ also shows a distinct $H$ dependence. Remarkably, $\ryx $ is almost flat up to $\mu _{0}H_{c}$=2.5 T.
The changes in both $\rxx $ and $\ryx $ appear to reflect the magnetic transitions. In fact, as shown in Fig. 1(g), the magnetization $M$ exhibits the similar step-like feature and seemingly reaches the expected value for 3/1 state ($\sim $2.68 $\mu _{\rm B}$/f.u. \cite{2005JPCSMachida}), although broadened at the slightly higher temperature (0.39 K).
More clearly, $dM/dH$ reveals two peaks at around 0.4 T and 2.3 T, each of which corresponds to the observed anomaly in respective transport properties. Therefore, we conclude that the Pr magnetic moments undergo the $H$-induced transitions from the ordered SI ($\mu _{0}H<$0.6 T) to KI which is the 2/2 state with polarized apical spins (0.6 T$<\mu _{0}H<$2.5 T), and finally to the fully-polarized 3/1 state (2.7 T$<\mu _{0}H$) as depicted on top of Fig. 1(c).
It is astounding because the emergence of both KI and SI is not immediately reproduced in the existing theory \cite{2003PRBMoessner}, as actually observed only in a few insulating materials such as Dy$_2$Ti$_2$O$_7$ \cite{2003PRLSakakibara}.
Nevertheless, the liquid-gas-like transitions similar to those in dipolar spin ice may occur in the ordered SI phase stabilized by higher-order RKKY interactions, if the long-range RKKY interaction gives rise to effective attractive interactions between the magnetic monopoles \cite{2014JPSCPIshizuka}.
All these findings imply that the Pr local moments undergo a cascade of magnetic transitions even in the metallic system $\pio $, in which the quantum-mechanical RKKY interaction plays a vital role in Pr magnetism rather than classical dipolar interactions, and furthermore, have a strong impact on the charge transport via possible topological transitions of the semimetal state.

\begin{figure}
\begin{center}
\includegraphics[width=2.5in,keepaspectratio=true]{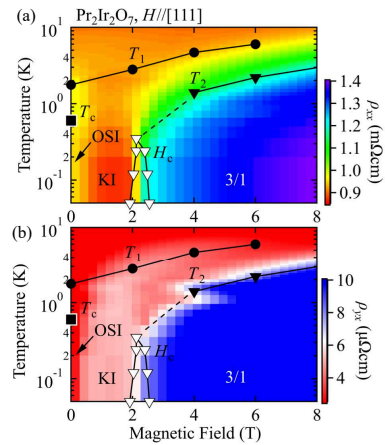}
\caption{(color online).
Contour plot of (a) resistivity and (b) Hall resistivity, respectively.
The filled square indicates the transition temperature $T_{\rm c}$ at zero field. The filled circles (triangles) denote the temperature $T_1$ ($T_2$) determined from Fig. 2(e), and the open triangle denote the critical magnetic field $H_{\rm c}$ determined from Figs. 1(e) and 1(f). OSI stands for ordered spin ice, KI stands for kagom$\rm \acute{e}$ ice, and 3/1 stands for 3-in 1-out state, respectively.
}
\end{center}
\end{figure}

Figures 2(a) and 2(b) show $H$ dependence of $\rxx $ and $\ryx $ at several temperatures, respectively.
The sharp features of $\rxx $ and $\ryx $ observed at 0.05 K diminish gradually as the temperature is elevated.
Nevertheless, the unique $H$ dependence is discernible up to 6 K well beyond $T_{\rm c}$. On the other hand, $M$ changes monotonically as functions of temperature and $H$ as shown in Fig. 2(c).
To gain more insight into the thermal evolution of the magnetic state, we measured the temperature dependence of the specific heat at several $H$ as shown in Fig. 2(d).
As $H$ increases, the sharp peak below $T_{\rm c}$ swiftly disappears at a small field of 0.3 T (see Ref. \cite{SM}) and other hump structures show up at around 1.4 K and 5 K at 4 T.
With further increasing $H$ up to 14 T, the humps progressively shift towards higher temperatures, extending the tail up to $\sim $30 K. To extract the magnetic contribution $\Delta C_{\rm m}$, we subtract the nuclear Schottky anomaly (denoted by the thick yellow line in Fig. 2(d)) and the lattice contribution estimated by the specific heat of the non-magnetic rare-earth material Eu$_2$Ir$_2$O$_7$ (for details, see Ref. \cite{SM}).
Figure 2(e) summarizes the temperature dependence of $\Delta C_{\rm m}$ at several $H$. A broad peak is centered at $T_1$=2 K and 0 T which possibly reflects the strong spin ice correlation, as predicted by Melko et al. \cite{2001PRLMelko}. The peak gradually drifts towards higher temperatures and becomes less pronounced with increasing $H$ as denoted by circles in Fig. 2(e).
An additional shoulder-like feature emerges at the lower temperature $T_2$ above 4 T, which slightly shifts to higher temperature as $H$ increases. The observed multiple peaks in $\Delta C_{\rm m}$ can stem from the Schottky anomalies of Zeeman-split spins \cite{2003JPSJHiroi}. $T_1$ indicates the freezing of the apical spins, i.e. the KI state, due to their larger Zeeman energy gain in the field along the anisotropic axis. On the other hand, the fully-polarized 3/1 state can be realized below $T_2$ at which the total Zeeman energy overcomes the energy barrier of the ice rule.

We show a contour plot of $\rxx $ and $\ryx $ in Figs. 3(a) and 3(b), combined with $T_1$ and $T_2$ determined by specific heat (filled circles and triangles) as well as $H_{\rm c}$ at which the hysteresis of $\rxx $ and $\ryx $ terminates (open triangles). As can be seen, both $\rxx $ and $\ryx $ are overall correlated with the magnetic transition points. For instance, both $\rxx $ and $\ryx $ are enhanced in the high field region below $T_2$, namely at the 3/1 state, indicative of the modulated electronic structure of WSM \cite{2017NCommUeda}.
Below $T_1$ and 2 T, i.e. the KI phase, $\rxx $ is small while $\ryx $ is relatively large as shown in Fig. 3(b).
On the other hand, right above 2.3 T and 0.4 K, where the critical end point terminates the first-order phase transition between the KI and 3/1 phases, $\rxx $ is relatively large as denoted in light green color, while $\ryx $ is small as denoted in red. This can be due to the strong fluctuation inherent to the precedence of the long-range order near the critical point \cite{2006PRLTabata,2007NPhysFennell}.

\begin{figure}
\begin{center}
\includegraphics[width=3.0in,keepaspectratio=true]{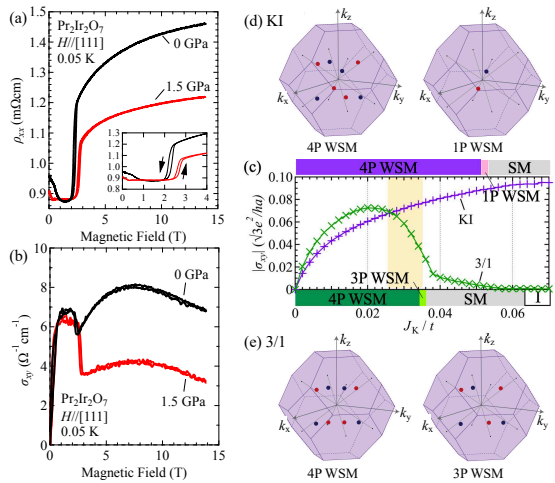}
\caption{(color online).
Magnetic field dependence of (a) resistivity and (b) Hall conductivity at the ambient pressure and 1.5 GPa.
(c) Theoretical calculation of Hall conductivity as a function of Kondo coupling $J_{\rm K}$. The green (purple) marks denote the Hall conductivity of 3-in 1-out (kagom$\rm \acute{e}$ ice) state in the unit of $\sqrt{3}e^{2}/ha$ ($a$ being the lattice constant). The top (bottom) figures show the schematic picture for the distribution of Weyl points for respective electronic phases.
}
\end{center}
\end{figure}

Depending on the $R$ (=Nd or Pr) moment order on the pyrochlore sublattice, $R_2$Ir$_2$O$_7$ is expected to show the versatile WSM state with various distributions of Weyl node pairs \cite{2018PRXYao,2018PRBOh,2017NCommUeda}. In this work, we have experimentally established all the possible 4$f$ moment orders, i.e. two types of 2/2 (SI and KI) orders as well as the 3/1 order, as a function of temperature and $H$. Each of the magnetic orders has its own WSM state, whose possible distribution of the Weyl node pairs in the momentum space is depicted in the upper and lower panels of Fig. 4(c) (for more detail of band structures, see Ref. \cite{SM}).
The variation of the magnetotransport presented here can be related to the topological transition among the WSM states induced by the temperature and/or $H$. To address this scenario, we employ the hydrostatic pressure which allows us to tune the physical parameters, such as the $f$-$d$ Kondo coupling, without introducing impurities or disorder. Figures 4(a) and 4(b) show $H$ dependence of $\rxx $ and Hall conductivity $\sxy $ at 0.05 K of our dilution refrigerator, when the sample is exposed to an ambient pressure and a hydrostatic pressure of 1.5 GPa.
As the pressure increases, the value of $\rxx $ remains almost unchanged in the intermediate field region (KI phase) while $H_{\rm c}$ obviously increases by 0.5 T, as shown in the inset to Fig. 4(a).
The transition between KI and 3/1 occurs under the competition between Zeeman energy and ferromagnetic interaction \cite{1998PRLHarris}.
Therefore, the increase of $H_{\rm c}$ means that the ferromagnetic RKKY interaction, which is proportional to the square of Kondo coupling $J_{\rm K} \propto t_{fd}^{2}/U$ ($t_{fd}$ and $U$ being $f$-$d$ transfer interaction and $f$-electron Coulomb repulsion, respectively), is increased by applying pressure.
For large $H$ regime, on the other hand, the pressure decreases $\rxx $ by nearly 15 \% at 14 T, suggesting that the electronic state of the 3/1 magnetic ordered phase is sensitive to the external perturbation. More importantly, $\sxy $ above 3 T significantly decreases nearly by half at 1.5 GPa, while $\sxy $ is almost intact below $H_{\rm c}$. This cannot be explained simply by the slight decrease of $\rxx $.

One plausible mechanism of the observed Hall effect is the modulation of electronic states endowed with Berry curvature. To investigate this scenario theoretically, we consider a Kondo lattice model with the Kondo coupling $J_{\rm K}$ between Ir $J$ =1/2 electron and Pr $4f$ moment, which is plugged into the Ir electronic system in pyrochlore lattice characterized by the QBT; the details of the theoretical calculation are described in Ref. \cite{SM}.
Figure 4(c) shows the calculated results of $J_{\rm K}/t$ ($t$ being Ir $d$-electron transfer interaction) dependence of $\sxy $ for KI and 3/1 states with schematics of respective electronic states for KI (Fig. 4(d)) and 3/1 (Fig. 4(e)), respectively. $\sxy $ for both KI and 3/1 increases singularly around $J_{\rm K}/t$=0 as a manifest of QBT \cite{2013PRLMoon}. $\sxy $ for the 3/1 state shows a maximum at $J_{\rm K}/t \sim $0.02 and then rapidly decreases above it, whereas $\sxy $ keeps increasing for KI.
The distinct $J_{\rm K}$ dependence is ascribed to the displacement of Weyl nodes in the momentum space for the respective WSM states. For the 3/1 state, the QBT at $\Gamma $ ($k$=0) point splits into four pairs of Weyl nodes and moves approximately along the [111] axes. The pair along the field direction, which maintains the three-fold rotation symmetry, eventually meets with each other and annihilates at the L points, turning into a semimetal with three pairs of Weyl nodes (3P WSM described on the right side of Fig. 4e). With further increasing $J_{\rm K}/t$, the remaining pairs experience annihilations, followed by the emergence of a trivial semimetal and insulator phase. In the course of this transition with increasing $J_{\rm K}/t$, $\sxy $ shows a non-monotonic curve toward zero in the insulator phase. On the other hand, for the KI state, in which the net magnetization of the Pr moments is smaller than that for 3/1, Weyl nodes keep from the pair annihilation up to $J_{\rm K}/t \sim $0.05. Therefore, $\sxy $ keeps increasing with $J_{\rm K}/t$.

In experiment, the application of the hydrostatic pressure should change both $J_{\rm K}$ and $t$ values, either by changing the lattice constant or by changing the bond angle. Judging from the critical increase of the KI-to-3/1 transition field $H_{\rm c}$ with pressure (see the inset to Fig.4(a)), the RKKY interaction, and hence the $J_{\rm K}/t$, is enhanced by the hydrostatic pressure.  According to the present calculation shown in Fig.4(c), the increase of $J_{\rm K}/t$ can cause a rapid reduction of $\sxy $ for 3/1 when $J_{\rm K}/t\sim $0.035, whereas $\sxy $ remains at a similar value for the KI when $J_{\rm K}/t\sim $0.025. Namely, if the $J_{\rm K}/t$ value can increase from $\sim $0.025 to $\sim $0.035 (by $\sim $40 \%) with application of 1.5 GPa (as indicated by the yellow shadow in Fig. 4(c)), the large pressure-induced change of $\sxy $ observed only in the 3/1 WSM can be well explained.  Note that this change is quite large or pressure-sensitive, yet is roughly in accordance with the observed change ($\sim $20\%) of the $H_{\rm c}$ (see the inset to Fig. 4(a)) which also reflects the pressure-induced change of $J_{\rm K}/t$.
	
To summarize, we investigate the magnetotransport properties of the pyrochlore $\pio$ crystal, a material predicted to exhibit rich topological phases. Both the resistivity and Hall resistivity exhibit a plateau-like step and then a discontinuous jump at the critical magnetic fields, indicating a large impact of magnetism on the charge transport. 
Our theoretical calculation suggests that the $\pio$ under magnetic field is in a magnetic Weyl semimetal induced by the $f$-$d$ coupling.
The present experiment-theory combined study demonstrates that the observed magnetic-field- and pressure-induced variations of the Hall effect can arise from the different motions of Weyl points in the momentum space for the versatile WSM, highlighting a promising direction towards the engineering of emergent topological functions.

We are grateful to F. Kagawa, M. Hirschberger and N. Nagaosa for fruitful discussion. The synchrotron radiation experiments were performed at SPring-8 with the approval of the Japan Synchrotron Radiation Research Institute (JASRI) (Proposals No. 2021B1261). This research was supported by JSPS/MEXT Grant-in-Aid for Scientific Research(s) (grant nos.19K1464, 21K13871, JP19K14649, JP18H03676) and CREST, JST (grant no. JPMJCR16F1).

\end{document}